\documentclass[12pt,oneside]{amsart}

\newcommand{\I}{\mathds I}

\usepackage{epsfig}
\usepackage{times}
\usepackage{dsfont}
\usepackage{amsmath}
\usepackage[hypertex]{hyperref}
\title[Scalar-field spacetimes with general
self-interaction potentials]%
{The dynamical behaviour of homogeneous scalar-field spacetimes with
general self-interaction potentials}

\author[R.\ Giamb\`o ,\ F.\ Giannoni]{Roberto Giamb\`o, Fabio Giannoni}
\address{Dipartimento di Matematica e Informatica \hfill\break\indent
Universit\`a di Camerino\hfill\break\indent Italy}
\email{roberto.giambo@unicam.it, fabio.giannoni@unicam.it}

\author[G.\ Magli]{Giulio Magli}
\address{Dipartimento di Matematica \hfill\break\indent Politecnico di
Milano \hfill\break\indent Italy} \email{magli@mate.polimi.it}

\begin{document}
\begin{abstract}

The dynamics of homogeneous Robertson--Walker cosmological models with a
self-interacting scalar field source is examined here in full
generality, requiring only the scalar field potential to be bounded
from below and divergent when the field diverges. In this way we are
able to give a unified treatment of all the already studied cases -
such as positive potentials which exhibit asymptotically polynomial
or exponential behaviors - together with its extension to a much
wider set of physically sensible potentials. Since the set includes
potentials with negative inferior bound, we are able to give, in
particular, the analysis of the asymptotically anti De Sitter states
for such cosmologies.

\end{abstract}

\maketitle

\section{Introduction}\label{sec:intro}


Scalar fields have attracted a great deal of attention in cosmology,
since the discovery that such a field can act as an "effective"
cosmological constant in driving inflation \cite{nde}. From this
first break-trough, which aroused from the simplest conceivable
model, i.e. that of a non self-interacting field, the research on
scalar field spacetimes has been constantly growing, and quite more
general scenarios have been considered, such as scalar fields
coupled with perfect fluids, non-minimal couplings, and alternate
theories of gravity \cite{all,vc1,vc3,vc4,vc5,bojo,cap,gvc3}

A scalar field spacetime can be viewed, like other matter models
coupled with gravity such as perfect fluids, as a solution of the
Einstein field equations which depends on the choice of the equation
of state for the matter. In the case of the scalar field, the role
of the equation of state is played by the self-interaction potential
$V$, which is equal to zero in the "standard" inflationary solution.
For several reasons, however, it is difficult to believe that this
function vanishes. For instance, dimensional reduction of
fundamental theories to four dimensions typically gives rise to
self-interacting scalar fields with exponential potentials, coupled
to four-dimensional gravity. Therefore exponential potentials have
been widely considered in the literature (see \cite{russo} and
references therein) also with the aim of uncovering possible
large-scale observable effects (\cite{fosh,st1,st2,topo}). Of
course, however, at the present status of our knowledge the specific
functional form of $V$ is quite unsure and, as a consequence, it
would be optimal to classify the dynamical behavior of the
spacetimes in dependance of all the possible choices of the
potential function. Relevant attempts have been made also in this
more general direction. Actually, the application of dynamical
systems techniques has proven very useful and, due mainly to the
works \cite{fos,mir1,mir2,mir3,r1,r2,rvc3}, we know the dynamical
behavior of scalar field spacetimes for a wide class of {\it
non-negative} potentials; for such potentials therefore, as will be
explained below, the contribution of the present paper relies in a
simplification and completion of known results.

Non negativity of the potential means that the energy density of the
scalar field has a positive lower bound. However, in many issues and
especially when string theory comes into play, it becomes relevant
to inspect spacetimes in which the potential still has a lower bound
but this bound is negative. From the physical point of view, it
should be noted that, although care must be given to the fact that
local positivity of energy density might be violated in these
spacetimes, under certain somewhat general conditions potentials of
this kind generate solutions with positive total energy \cite{hert}.
In particular, of course, a constant negative potential generates an
"equilibrium" state which is just the Anti de Sitter solution (AdS);
therefore, a potential with a negative minimum generates a class of
spacetimes which have an AdS "equilibrium" (see
e.g. \cite{dafermos,hor,hor1}). In the present paper  we thus study the
dynamical behavior of FRW scalar field cosmological models, imposing
only very general conditions on the scalar field potential; our
hypotheses essentially reduce to ask the potential to be bounded
from below and divergent when the field diverges.  Within this quite
general framework, we give a unified treatment of all the relevant
known cases - such as asymptotically polynomial and exponential
behaviors - as well as its completion to a much wider set of
possible, physically sensible potentials, also in presence of
negative (i.e. AdS) minima. It is worth mentioning that our
treatment can be applied to the "reverse" problem as well, i.e. to
homogeneous scalar field collapse. This issue is treated in
companion papers \cite{hsf,jmp,collapse}.

The paper is organized as follows. First, we formulate the field
equations as a first order, regular dynamical system for the scalar
field, its time derivative, and the scale factor. Then, we study in
full generality the behavior of the trajectories of this system in
dependance of the properties of the potential function. Finally, the
issue of recollapse and that of stability of AdS spacetimes are
addressed. In particular, we show that that anti deSitter solutions are the only solutions such that $\phi$ goes to a critical point of $V(\phi)$ with negative critical value.

\section{Formulation of the problem}

Einstein's field equations for the Robertson--Walker model
\begin{equation}\label{eq:FRW}
\mathrm ds^2=-dt^2 + a(t)^2\left[\frac1{1-k r^2}\,\mathrm
dr^2+r^2\left(\mathrm d\theta^2+\sin^2\theta\,\mathrm
d\varphi^2\right)\right],
\end{equation}
coupled with the stress--energy tensor
\begin{equation}\label{eq:Tgen}
4\pi T_{\mu\nu}=\partial_\mu\phi\partial_\nu\phi-\left(\frac12
g^{\alpha\beta}\partial_\alpha\phi\partial_\beta\phi+V(\phi)\right)g_{\mu\nu
},
\end{equation}
are given by
\begin{equation}\label{eq:G00}
(G^0_0=8\pi T^0_0):\qquad-\frac{3(k+\dot
a^2)}{a^2}=-(\dot\phi^2+2V(\phi)),
\end{equation}
\begin{equation}\label{eq:G11}
(G^1_1=8\pi T^1_1):\qquad-\frac{(k+\dot a^2)+2a\ddot
a}{a^2}=(\dot\phi^2-2V(\phi)).
\end{equation}
in the unknown functions $a(t),\phi(t)$. These equations imply
\begin{equation}\label{eq:KG}
T^\mu_{\,\,0;\mu}=-2\dot\phi\left(\ddot\phi+V'(\phi)+3\frac{\dot
a}a\dot\phi\right)=0.
\end{equation}
We transform the above second order system
\eqref{eq:G00}--\eqref{eq:G11} into a first order system introducing
the additional variables $h=\tfrac{\dot a}a, v=\dot\phi$. We must
exclude solutions such that $\phi(t)=\phi_0$ constant on some
interval but $V'(\phi_0)\ne 0$\footnote{these solutions are easily
found by integration of \eqref{eq:G00} that becomes $\dot
a^2=\tfrac{2V(\phi_0)}3a^2-k$.}, so that we are sure that
\eqref{eq:KG} holds due to the vanishing of the term in brackets.
Therefore $(\phi,v,h)$ solves the regular system
\begin{align}
\dot\phi&=v,\label{eq:dphi}\\
\dot v&=-V'(\phi)-3h\,v\,\label{eq:dv}\\
\dot h&=-h^2-\tfrac23(v^2-V(\phi)).\label{eq:dh}
\end{align}
To see that the converse also holds true, we define the function
\begin{equation}\label{eq:W}
W(\phi,v,h)=h^2-\frac13(v^2+2V(\phi)),
\end{equation}
and argue as follows. Given a solution $(\phi(t),v(t),h(t))$ of
\eqref{eq:dphi}--\eqref{eq:dh}, then $\dot W(t)$, the derivative of
$W(t)$ along the solution, satisfies the differential equation
\begin{equation}\label{eq:dotW}
\dot W=-2hW,
\end{equation}
that yields $W(t)=W_0 \exp(-2\int_0^t h(\tau)\mathrm d\tau)$.
Moreover, we set
\begin{equation}\label{eq:a}
a(t)=a_0 e^{\int_0^t h(\tau)\mathrm d\tau},
\end{equation}
where $a_0=|W_0|^{-1/2}$ if $W_0\ne 0$, and $a(0)=1$ if $W_0=0$.
Then the identity
\begin{equation}\label{eq:Wk}
W(t)=-\frac{k}{a(t)^2}
\end{equation}
holds, where $k=-\mathrm{sgn}(W_0)$, and so $\phi(t)$ and $a(t)$
solve \eqref{eq:G00}--\eqref{eq:KG}, from which \eqref{eq:G11}
follows.

We stress some features of the above approach: first, the system
obtained is regular, with equilibrium points given by
$(\phi_*,0,h_*)$ such that $V'(\phi_*)=0,\,h_*^2=\frac23 V(\phi_*)$
(notice that $W(\phi_*,0,h_*)=0$), that
physically correspond to deSitter universe.
Given a solution $(\phi(t),v(t),h(t))$, then $(\phi(-t),-v(-t),-h(-t))$ is the time reversed solution.
Moreover, the system represents
solutions of \emph{all} Robertson--Walker cosmologies: indeed, from
\eqref{eq:dphi}, the set $W^{-1}(0)$ is invariant by the flow, and
by local uniqueness this ensures that the sign of $W$ is invariant
along the flow; from \eqref{eq:Wk} we deduce that solutions with $W$
positive, null, or negative, represent scalar field cosmologies with
$k=-1,0,1$ respectively. From the sign of $h(t)$ one can tell
whether, at the instant $t$, the solution is collapsing ($h<0$) or
expanding ($h>0$).

\section{Qualitative analysis of the expanding models}

In this section we consider initially expanding spacetimes, i.e.
solutions such that $h(0)=h_0>0$. We suppose that the potential
$V(\phi)$ is a $C^2$ function, such that
$\lim_{\vert\phi\vert\to\infty} V(\phi)=+\infty$. We will suppose
that critical points of $V$ are isolated, and they are either
minimum points or nondegenerate maximum points. We also impose the
weak energy condition to be satisfied at initial time, i.e.
$v_0^2+2V(\phi_0)>0$. Of course, this is not sufficient for the
w.e.c. to be satisfied along the evolution, since $V(\phi)$ can
attain also negative values. It is useful to introduce the energy
function $$\epsilon=v^2+2V(\phi),$$ in such a way that
\eqref{eq:dphi}--\eqref{eq:dh} imply
\begin{equation}\label{eq:de}
\dot\epsilon=-3hv^2.
\end{equation}
If we let the solution evolve, either there exists $T>0$ such that
$h(T)=0$, or $h(t)>0$ for all $t<\sup\I$, where $\I$ is the maximal
interval of definition of the solution. In the latter case, from
\eqref{eq:dotW} we deduce that $|W(t)|\le|W_0|$ and moreover,
recalling \eqref{eq:de}, $\epsilon\le \epsilon_0=v_0^2+2V(\phi_0)$.
Notice also that $\epsilon(t)$ is bounded from below since $V(\phi)$
is, and this implies that both $\phi(t)$ and $v(t)$ are bounded.

Supposing $h(t)$ always positive during the evolution, let us
consider separately the cases $W\ge 0$ and $W<0$. In the first case
$h$ decreases, since \eqref{eq:dh} and \eqref{eq:W} imply
\begin{equation}\label{eq:dh2}
\dot h=-W-v^2,
\end{equation}
and so $h$ also is bounded. On the other side, $W<0$ implies
$h^2<\tfrac\epsilon 3$ and so $h$ is bounded again. In any case the
solution lives in a compact set, and this means that
$\sup\I=+\infty$, and it is easily proven that it converges to an
equilibrium point $(\phi_*,0,h_*)$ of the system.

Translated into the original formulation, we deduce that if an
originally expanding solution does not recollapse, then it is
regular for all times, the scalar $\phi$ converges to a  critical
point $\phi_*$ of the potential $V$, with nonnegative critical
value, and $\dot\phi\to 0$.

\begin{figure}
\begin{center}
\psfull \epsfig{file=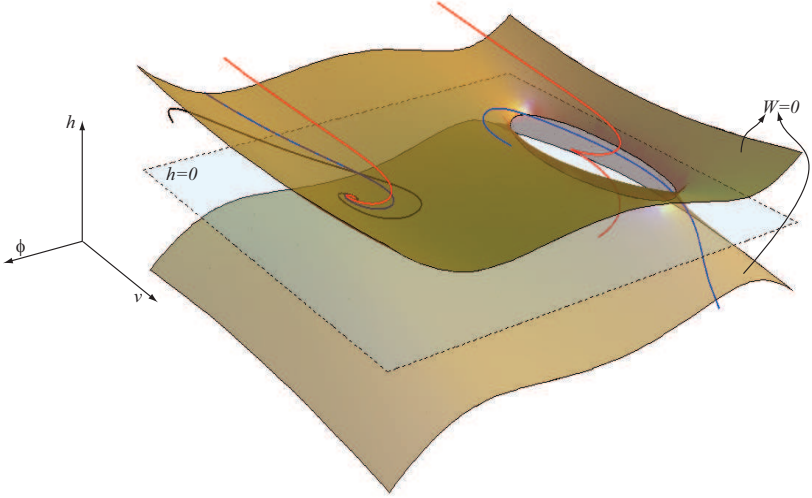, height=9cm}
\caption{Phase portrait sketch when $V(\phi)=5 - 6 x^2 + x^3 + 2 x^4$. The behavior when $t>0$ for some curves with initial data in the open half--space $\{h>0\}$
is represented. On the left, the situation nearby a local minimum  $\phi_*$ for $V$, with positive critical value. The three curves on the left side, living in the regions where $W$ is positive, negative and null, respectively, all converge to the equilibrium point $(\phi_*,0,h_*)$, and therefore the corresponding cosmologies are expanding forever. On the right, the situation nearby a local minimum of $V$ with negative critical value. The two curves, such that $W$ is positive and null, respectively, leave the half space $\{h>0\}$,  therefore passing from expansion to collapsing phase.
}\label{fig:dyna2}
\end{center}
\end{figure}
An example is sketched in Figure \ref{fig:dyna2}, when the phase portrait for the system \eqref{eq:dphi}--\eqref{eq:dh} is sketched, for a potential that has two local minima, one with positive critical value, the other one with negative critical value. In the first case, every solution starting nearby the critical point $(\phi_*,0,\sqrt{2V(\phi_*)/3})$ approaches it, regardless of the curvature ($k=-1,0$, or 1). In the second case, the negative minimum determines an intersection between the two branches of the set $W=0$, that acts as a ``tunnel'' allowing solution such that $W\ge 0$ and starting with $h>0$ to recollapse.

The recollapse problem is an interesting feature of these
cosmological models that deserves to be further investigated (see
Section \ref{sec:rec}). At the moment, we observe that \eqref{eq:W}
-- but a simple inspection of \eqref{eq:G00} also -- implies that
solutions with $k=-1$ may recollapse only if they violate the
w.e.c., and, by the aforesaid, that occurs only if $V$ can attain
negative values. We will turn back on this situation later when we
will examine stability of anti de Sitter model.

Now we examine stability of the equilibria  $(\phi_*,0,h_*)$ as
expanding solutions, where $\phi_*$ is a critical point of the
functional and $h_*=\sqrt{\frac{2V(\phi_*)}3}$. Recall that, by the assumptions made on $V(\phi)$, $\phi_*$ is either a minimum, or a non degenerate maximum.
Let us now briefly review the second situation: here, one can simply
linearize the system \eqref{eq:dphi}--\eqref{eq:dh} to find that
eigenvalues are always real, and one of them is always positive, and
one always negative, which results in the existence of a 1- or
2-dimensional unstable manifold at the equilibrium point, and then, up to a zero--measure set of initial data, the scalar field do not approach the maximum.

Now,
consider the case when $\phi_*$ is local minimum for $V$,
with non negative critical value. If $\phi_*$ is a non degenerate
critical value, asymptotic stability of the equilibrium is found simply using the linearized flow of \eqref{eq:dphi}--\eqref{eq:dh}, and therefore solutions with initial data near the equilibrium point approach it, regardless the case $k$ under consideration.
Here, we want to extend to the case $\phi_*$ is possibly degenerate.

Let us begin considering the situation $V(\phi_*)\ge 0$, in the open
topologies $W\ge 0$. Let us consider first the case $V(\phi_*)$ is
\emph{strictly} positive. We call $\bar V>V(\phi_*)$ a regular value
such that $\phi_*$ is the only critical point in the connected
component of $V^{-1}(]-\infty,\bar V])$ containing $\phi_*$, and let
us consider a solution $(\phi(t),v(t),h(t))$ with initial data such
that $v_0^2+2V(\phi_0)\le 2\bar V$, and let $\bar W=W(0)>0$. We define
$\Omega$ to be the connected component of the set
$\{(\phi,v,h)\,:\,W(\phi,v,h)\in[0,\bar
W],\,\epsilon(\phi,v)\in[2V(\phi_*),2\bar V]\}$ containing the
equilibrium point $(\phi_*,0,\sqrt{\frac{2V(\phi_*)}3})$. It is easy
to see, using \eqref{eq:dotW} and \eqref{eq:de}, that $\Omega$ is
compact, and positively invariant by the flow. So, let us consider a solution
with initial data in $\Omega$. Applying LaSalle's invariance theorem
\cite{wig} to the functions $W$ and $\epsilon$, we see that the
solution must be such that $hW\to 0$ and $h v^2\to 0$.
But $h$ is strictly bounded away from zero in $\Omega$, then both
$W$ and $v$ must go to zero, which means that the solution
approaches the equilibrium point. Since the critical value is strictly
positive, then \eqref{eq:a} says that the scale factor $a(t)$
diverges at infinity like $\exp(t\sqrt{\tfrac23V(\phi_*)})$.

\begin{figure}
\begin{center}
\psfull \epsfig{file=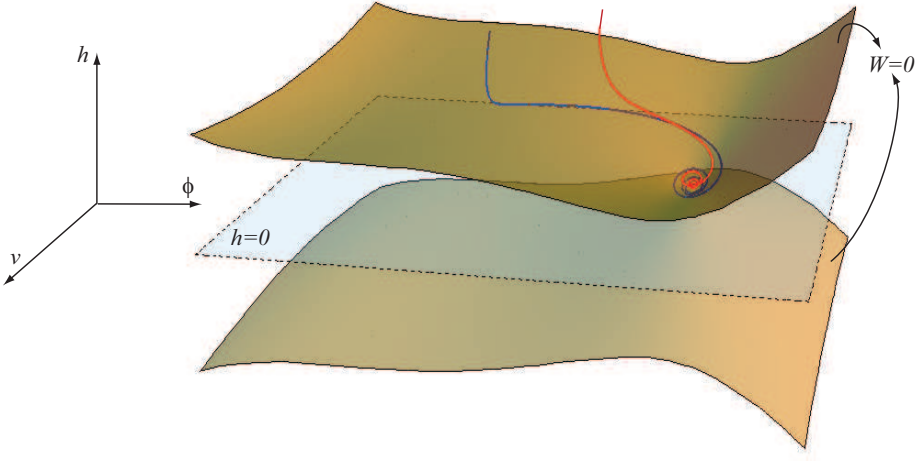, height=9cm}
\caption{Phase portrait sketch when $V(\phi)=x^2 + (2 x^3)/3 + x^4/8$. The behavior of two curves with initial data in the half space $\{h>0\}$, living in the regions where $W$ is positive and null, respectively, is represented. The two curves both approach the equilibrium point $(0,0,0)$, where $\phi_*=0$ is the local minimum of $V$, with zero critical value. Both curves spin around the equilibrium point for late times.}\label{fig:dyna3}
\end{center}
\end{figure}

If $V(\phi_*)=0$, the above argument can be easily adapted. In this
case the set $\{(\phi,v,h)\,:\,W(\phi,v,h)\in[0,\bar
W],\,\epsilon(\phi,v)\in[2V(\phi_*),2\bar V]\}$ is connected, and
we choose $\Omega$ to be its subset characterized by the property $h\ge
0$. The only point in $\Omega$ with $h=0$ is exactly the equilibrium
point, and so if $h(t)\to 0$ (recall $h$ is monotone) the solution
is forced to approach the equilibrium. If by contradiction
$h(t)$ had a strictly positive limit, we can argue as before to find
$v\to 0$ and $W\to 0$ and so $h\to 0$. Moreover, when the minimum is non degenerate, then the
couple $(\phi,v)$ has an oscillatory behavior. This fact was already
pointed out by Rendall for the flat case \cite{r2}, but it actually
takes place also when $k=-1$. Indeed, taking $\phi_*=0$ for sake of simplicity, and using the variable change
$\phi=\lambda^{-1/2}r \cos\theta$,  $v=r\sin\theta$,
where $2V(\phi)=\lambda\phi^2+U(\phi)$, with $U(\phi)=o(\phi^2)$, then the equation for $\theta$ reads
$$
\dot\theta=-\sqrt\lambda-\frac32 h\sin 2\theta-\frac{\cos\theta}{2r}U'(\lambda^{-1/2}r\cos\theta),
$$
and since both $r$ and $h$ go to zero, we deduce that $\theta(t)=-\sqrt\lambda t+o(1/t)$, and this implies the oscillatory behavior of $\phi$ (and $v$). An example of this is sketched in Figure \ref{fig:dyna3}

Then, it remains proven that every solution of flat or negatively
curved Robertson--Walker cosmology with initial data near the local
minimum of the potential expands forever approaching this minimum,
and the velocity of the scalar field vanishes at infinity.

If $W<0$, we must adapt the above argument, since now we do not have a priori information on monotony of $h$ (see \eqref{eq:dh2}).
As before, we start from the case $V(\phi_*)>0$, and take $\bar
V$ and initial data such that $v_0^2+2V(\phi_0)\le2\bar V$, whereas
now $\bar W$ will be a negative value to determine in such a way
that it acts as a lower bound for $W$. Supposing that initial data are taken near the equilibrium, then $h$ starts
positive; since $W(0)\ge \bar W$,
\eqref{eq:dotW}and \eqref{eq:de} imply that $W\ge \bar W$
and $\epsilon(t)\le 2\bar V$. Last relation, in particular, shows that $V(\phi(t)\le\bar V$ and this means that $\phi(t)$ is forced to stay in the potential well $V(\phi_*)\le V(\phi(t))\le\bar V$. Then
$$
h^2\ge \frac\epsilon 3+\bar W\ge\frac{2V(\phi_*)}3+\bar W=h_*^2+\bar
W,
$$
that is
$$
h\ge \left(\sqrt{1+\frac{\bar W}{h^2_*}}\right)h_*=:\bar h.
$$
Then, chosen $\bar W$ sufficiently small such that $h^2_*>-\bar W$,
we have that $h(t)>0\Rightarrow h(t)\ge \bar h>0$. Therefore, we can apply again LaSalle's invariance theorem as in the case $W\ge 0$ and exploit the fact that $h$ is bounded away from zero.

All in all,
a result of stability of local minima of $V(\phi)$ with positive
critical value holds also in the positively curved Robertson--Walker
cosmologies. This result, however,
cannot be extended as before to the case $V(\phi_*)=0$, because the
equilibrium is such that
$h_*=0$, and so nearby solutions may recollapse
(i.e. $h$ can change sign).

\section{Recollapsing solutions and the instability of anti deSitter
models}\label{sec:rec}

In this section we give some answer to the
recollapsing solutions problem sketched before, specifically in the open topology cosmologies, and  address the
question of stability of anti de Sitter solutions with respect to
perturbations  in the Robertson--Walker cosmologies.

Let us consider a solution of \eqref{eq:dphi}--\eqref{eq:dh} with
$h(0)>0$. Since the sign of $W$ is invariant along the flow, we know
that $W(0)\ge 0$ implies, using  \eqref{eq:dh}, that $h(t)$ is
monotone decreasing. If the potential $V(\phi)$ is nonnegative, we
already know from previous section that no recollapse will take
place, and the solution will expand for all times. So, let $\bar V$
be such that $V^{-1}(]-\infty,\bar V])$ has a connected component
containing only a critical point $\phi_m$ which is a local minimum
with negative critical value. Taking initial data such that
$v_0^2+2V(\phi_0)\le 2\bar V$, then $\epsilon$ and $W$ will decrease
until $h$ remains positive, and the solution is forced to remain in
the (compact) set such that $W\in[0,W(0)]$, $\epsilon\le 2\bar V$.
But  there are no equilibrium points to approach in this set,
which means that there must be a time $T$ when $h(T)=0$, and the
solution recollapse (notice that this happens violating w.e.c.).

Among these solutions, we can consider anti deSitter models, which
occur when $\phi(t)\equiv \phi_m$, and $v(t)=0$.  Let us call
$\mu^2=-\tfrac23 V(\phi_m)$; then \eqref{eq:dh} takes the form $\dot
h=-h^2-\mu^2$ that integrates to give $h(t)=-\mu\tan(\mu(t-c))$,
where $c$ is the constant of integration. Then $h(t)$ diverges to
$-\infty$ in a finite time $t_s>0$, and so by \eqref{eq:W}
$W\to+\infty$, and by \eqref{eq:Wk}, $a(t)$ goes to zero, but this
is actually a ``false'' singularity, that does not correspond to a
divergence of stress energy tensor which instead remains finite,
since $v$ and $\phi$ are constant.

In order to investigate stability of these solutions, we choose
initial data $(\phi_0,v_0)$ near $(\phi_m,0)$, and suppose that
$(\phi(t),v(t),h(t))$ is a solution of
\eqref{eq:dphi}--\eqref{eq:de}, defined in an interval $\I$, such
that $\phi(t)\to\phi_m$, $v(t)\to 0$. Then $h(t)$ diverges to
$-\infty$ as $t\to \sup\I$.
Indeed, by \eqref{eq:dh}, and recalling $V(\phi_m)<0$,  $h(t)$ is eventually decreasing, but if $h(t)$ admits a finite value it would be bounded, so that $\sup\I=+\infty$ and $\dot h(t)$ would tend to a negative value, which cannot happen.
This fact implies, again, that $a(t)$ goes to
zero, that means, using \eqref{eq:a}, that
$$\lim_{t\to\sup\I}\int_0^{t}h(\tau)\,\mathrm d\tau=-\infty.$$
Now let us consider equation \eqref{eq:dv}, and call $f(t)$ the
(bounded) function $-V'(\phi(t))$. Then it must be
$$
\dot v(t)+3h(t) v(t)=f(t),
$$
which can be integrated to give a diverging solution, once we prove
that $v_0\ne 0$. Indeed, one can consider situations where $v_0=0$,
but if $v(t)=0$ for all time, \eqref{eq:dv} would give
$V'(\phi(t))=0$ for all time, that means $\phi(t)=\phi_m$, obtaining
anti deSitter model once again. Then $v_0$ can be taken nonzero
without loss of generality, and this results in a diverging solution
$v(t)$, which is in contradiction with the assumption $\phi,v$
bounded. We conclude that anti deSitter solutions are the only solutions of \eqref{eq:dphi}--\eqref{eq:dh} such that $\phi$ goes to a critical point of $V(\phi)$ with negative critical value.

\section{Discussion and conclusions}\label{sec:intro}

We analysed here the dynamics of homogeneous Robertson--Walker
cosmological models with a self-interacting scalar field source
within a quite general class of (physically valid) potentials
$V(\phi)$, essentially requiring only the function $V$  to be
bounded from below and divergent when the field diverges. The
analysis has been carried out by casting the problem as a regular
dynamical system whose solutions describe the possible Robertson
Walker cosmologies together with the scalar field evolution.

The asymptotic behavior of the expanding (cosmological) solutions
depends crucially on the sign and the extrema of the function $V$.
First of all, if $V$ is positive, the scalar field always approaches
the minimum and the solutions do never re-collapse. This result
extends and completes already existing results, for instance for
exponential potentials, and holds also for the special case of
vanishing $V$ at the minimum if the topology is not closed.

The situation complicates drastically for potentials with a negative
lower bound. Indeed, solutions which are close to the negative
minimum are repulsed; from the cosmological point of view, the
corresponding universes re-collapse. The unique exception is the
solutions which "sits" on the negative minimum, which of course is
the Anti De Sitter space-time. In this sense, AdS turns out to be
unstable within homogeneous scalar fields cosmologies.

The phase space in case of several extrema can be sketched using as
an example a potential with a positive maximum and one negative and
one positive minimum. Clearly, from the results above, the phase
space contains ever expanding solutions approaching the positive
minimum, and recollapsing solutions. The two sets are separated by
solutions for which the scalar field tends to the maximum of the
potential. It can be easily shown however that the initial data
leading to such a situation form a two-dimensional surface, and
therefore these data, besides being of course unstable, are not
generic.

\end{document}